\documentclass[aip,jcp,amsmath,amssymb,reprint,longbibliography,groupedaddress]{revtex4-1}
\usepackage{amsmath}
\usepackage{braket}
\usepackage[utf8]{inputenc}
\usepackage{graphicx}
\usepackage{color}
\usepackage{multirow}
\usepackage{comment}
\usepackage{subcaption}
\usepackage[Euler]{upgreek}
\usepackage{makecell}
\usepackage{textgreek}
\usepackage{bm}
\usepackage{hyperref}
\DeclareMathOperator{\Tr}{Tr}

\newcommand{\iu}{\ensuremath{\mathrm{i}}}
\newcommand{\eu}{\mathrm{e}^}

\newcommand{\rmd}{\mathrm{d}}
\newcommand{\half}{{\ensuremath{\frac{1}{2}}}}

\newcommand{\thalf}{{\ensuremath{\tfrac{1}{2}}}} 

\newcommand{\eqn}[1]{Eq.~\eqref{#1}}

\newcommand{\pder}[3][]{\frac{\partial^{#1}{#2}}{\partial{#3}^{#1}}}
\newcommand{\pders}[3]{\frac{\partial^2{#1}}{\partial{#2}\partial{#3}}}
\usepackage{dcolumn} 
\newcolumntype{d}[1]{D{.}{.}{#1}} 

\newcommand{\op}[1]{\hat{#1}}

\usepackage{soul} 

\begin{document}

\title{Ring-polymer instanton theory for tunneling between asymmetric wells}
\author{Marit R. Fiechter}
\altaffiliation{These authors contributed equally}
\author{Gabriel Laude}
\altaffiliation{These authors contributed equally}
\author{Jeremy O. Richardson}
\email{jeremy.richardson@phys.chem.ethz.ch}
\affiliation{\mbox{Institute of Molecular Physical Science, ETH Z\"urich, 8093 Z\"urich, Switzerland}}

\begin{abstract} 
    Instanton theory has arisen as a practical tool for calculating tunneling splittings in molecular systems. Unfortunately, the original formulation of instanton theory fundamentally breaks down when trying to calculate the level splitting in asymmetric double wells, as there is no imaginary-time periodic orbit connecting the two non-degenerate minima.
    We have therefore developed a new formulation of instanton theory based on a projected flux correlation function that is applicable to these asymmetric systems. 
    Comparison with exact quantum-mechanical results in one- and two-dimensional models demonstrates that it has a reasonably high accuracy, similar to that reported for instanton theory in the symmetric case.
    The theory is then applied to study tunneling between non-degenerate minima in the biomolecule \textalpha-fenchol,
    for which we find good agreement with experiment.
    Finally, we use the connection to instanton rate theory, which is also derived from flux correlation functions, to discuss the often misunderstood 
    relationship between tunneling splittings and reaction rate constants.
\end{abstract}

\maketitle


\section{Introduction}  


Since the early days of quantum mechanics, it has been well known that tunneling in a symmetric double-well potential results in a splitting of the energy levels.
\cite{hund1927deutung,dennison1932two,dennison1932parallel}
In contrast, few studies have been conducted on molecules with asymmetric double wells,
for which tunneling may also play an important role if the asymmetry is not too strong.
\cite{rossetti1981proton,fenchol}
Such systems may be realized by molecules in which a proton transfers between two similar but not identical functional groups,
or for symmetric tunneling in molecules embedded in a weakly-interacting asymmetric environment.
New methodology is required for first-principles simulations of these interesting cases.

Many theoretical methods have been developed to calculate tunneling splittings in symmetric molecular systems.
Some approaches aim to compute the exact quantum-mechanical result and can be categorized into truncated basis-set methods,
\cite{Hammer2011malonaldehyde,Schroeder2011malonaldehyde,wu2016hydrogen,Lauvergnat2023malonaldehyde,Wang2018dimer,Simko2025trimer} 
diffusion Monte Carlo (DMC),
\cite{sun1990diffusion,gregory1995calculations,gregory1995three,gregory1996tunneling,liu1996characterization,viel2007ground,wang2008,qu2021full} 
and path-integral methods. \cite{Ceperley1987exchange,Alexandrou1988tunnelling,Marchi1991tunnelling,Benjamin2005PIMD,Matyus2016tunnel1,*Matyus2016tunnel2,Vaillant2018dimer,*Vaillant2019water,Zhu2022trimer,PIMDtunnel,malonaldehydePIMD}
Due to the difficulty of converging these simulations, especially in high-dimensional problems,
it is often necessary
to use reduced dimensionality,
\cite{bosch1990bidimensional}
reaction path/surface Hamiltonians,
\cite{carrington1986reaction,shida1989theoretical,Wang2008malonaldehydeSplit}
assume a decoupling between inter- and intramolecular degrees of freedom,
\cite{althorpe1994calculation,althorpe1995new,fellers1999fully,smit2001vibrations,huang2008new,Felker2024trimer}
or to introduce semiclassical approximations,
such as WKB theory, \cite{landau1965lifshitz,garg2000tunnel}
periodic-orbit theory \cite{Miller1979periodic}
or Bohr--Sommerfeld quantization. \cite{chebotarev1998extensions,child2014semiclassical}
Although suggestions have been made to apply these semiclassical methods along the tunneling path of a full-dimensional molecule,
\cite{makri1989semiclassical,sewell1995semiclassical,Wales1993tunnel,tautermann2002optimal}
they are formally derived only for one-dimensional systems.



A rigorous multidimensional semiclassical approach is provided by instanton theory.
\cite{Miller1971density,Uses_of_Instantons,ABCofInstantons,Benderskii,Kleinert,mil2001practical,tunnel,InstReview,Erakovic2020instanton} 
It is derived from a semiclassical approximation to the path-integral formulation of quantum mechanics \cite{Feynman}
and is based on a uniquely-defined optimal tunneling path (called the ``instanton'').
The instanton path is defined via a stationary-action condition, which makes it equivalent to an imaginary-time trajectory, or, put another way, a classical trajectory obeying Newton's equations of motion for the upside-down potential.\cite{Miller1971density}
In the symmetric case, the instanton path is a periodic orbit which connects the bottoms of the two wells.
The ring-polymer instanton (RPI) method has become a popular tool to calculate ground-state tunneling splittings in molecular systems. \cite{Perspective,water,octamer,hexamerprism,Cvitas2016instanton,Cvitas2018instanton,vaillant2018rotation,formic,i-wat2,sahu2021instanton,tropolone,TransferLearning,TransferLearning2,Videla2023tunnel}
Apart from providing an intuitive picture of the dominant tunneling path, its main advantages over other methods is that
instanton theory can treat complex molecular systems in full dimensionality with only modest computational resources.
%
In practice, the full-dimensional semiclassical instanton approximation is often more accurate than a exact solution of the Schr\"odinger equation for a reduced-dimensionality model. \cite{formic}


Some of the standard methods can be applied directly to asymmetric double wells, \cite{Bosch1992malonaldehyde,Miller1979periodic,sorenson1996diffusion,song2015localization, Halataei2017WKB, pollak2025nonadiabaticsplittings}
but the original derivations of instanton theory are applicable only to symmetric systems. \cite{Uses_of_Instantons,Benderskii,Kleinert,tunnel}
Recently, we generalized instanton theory for molecules with asymmetric isotopic substitutions such that the wells are equally deep but have different widths and therefore different zero-point energies (ZPEs). \cite{asymtunnel} The energy levels in this case are split due to a combination of asymmetry and tunneling.
In this work, we go even further and develop an instanton theory that can be applied to the more general type of asymmetric double-well system in which the wells are of unequal depth as well as having different ZPEs.
In this case, when one carries out a semiclassical analysis of the partition function, there are no minimum-action pathways which contribute other than the non-tunneling paths collapsed in each of the two wells.
The reason for this is that there cannot be an imaginary-time classical periodic orbit which connects the bottoms of the two wells if the wells have different potential energies.\footnote{Although no periodic orbit exists, there is an imaginary-time trajectory which runs from one minimum to the other.  One might try to construct an asymmetric instanton theory based on this trajectory, but this is complicated by the fact that the zero-frequency permutational mode only appears when the two minima have exactly the same energy.}


Therefore, rather than following our previous derivation based on a semiclassical analysis of the partition function,
our new approach for calculating the level splitting will start from the projected flux correlation function introduced by Ref.~\onlinecite{QInst} in a different context.
As shown in Sec.~\ref{sec:theory}, following this approach allows for the definition of an instanton pathway in asymmetric double wells.  This generalized expression for the tunneling frequency reduces to the original formulation in the symmetric case.
We demonstrate the accuracy of the method by comparison to numerically exact quantum benchmarks in one- and two-dimensional models, and perform a full-dimensional first-principles calculation of the level splitting in the biomolecule \textalpha-fenchol in Sec.~\ref{sec:method}\@.
Finally, since the flux correlation function also forms the basis for instanton rate theory, \cite{QInst,AdiabaticGreens,InstReview,Lawrence2024uniform} this new formulation allows us to study the relation between the tunneling splitting and the low-temperature limit of the reaction rate. As we show in Sec.~\ref{subsec:discuss-rate}, this analysis contradicts a simple approximation often quoted in the literature.  
We conclude the article in Sec.~\ref{sec:conclusions}, where we also discuss an alternative asymmetric instanton method developed by Erakovi\'c and Cvita\v{s} from a completely different starting point.\cite{Erakovic2022instanton}

\section{Theory} \label{sec:theory}

The standard derivation of instanton theory starts by considering a semiclassical approximation to the partition function of a symmetric double-well system. \cite{Benderskii,tunnel}
By comparing this to the algebraic result for an effective two-level Hamiltonian, an estimate of the tunneling splitting can be obtained.
As mentioned above, this approach cannot be applied to wells of unequal depth.
We therefore derive the theory based on a different quantity, 
namely the projected flux correlation function introduced in Ref.~\onlinecite{QInst} .
It generalizes the well-known flux correlation function \cite{Miller1983rate}
to a more powerful asymmetric form, which was originally designed to correct the quantum instanton method and lead more directly to instanton rate theory when the semiclassical limit is taken.
It is defined by
\begin{align} \label{eq:cff_definition}
    c_\text{ff}(\tau) &= \Tr\big[ \op{F} \op{K}_\ell(\tau) \op{F} \op{K}_r(\beta\hbar-\tau) \big] ,
\end{align}
where 
$\op{F}$ is the flux operator.
The definition of the projected imaginary-time propagators is quite flexible, but the simplest choice is
\begin{subequations}
\begin{align}
    \op{K}_\ell(\tau_\ell) &= \eu{-\tau_\ell\op{H}/2\hbar} \, \op{P}_\ell \, \eu{-\tau_\ell\op{H}/2\hbar} ,
    \\
    \op{K}_r(\tau_r) &= \eu{-\tau_r\op{H}/2\hbar} \, \op{P}_r \, \eu{-\tau_r\op{H}/2\hbar} ,
\end{align}
\end{subequations}
where $\op{H}$ is the Hamiltonian, which typically has the form
$\op{H}=\op{p}^2/2m+V(\op{x})$ for a particle of mass $m$ in a double-well potential, $V(x)$.
In one dimension, the projection operators are defined in terms of the Heaviside step function as
$\op{P}_\ell = \theta(x_\sigma-\op{x})$ and 
$\op{P}_r = \theta(\op{x}-x_\sigma)$, where $x_\sigma$ is the position of the dividing surface, typically located at the barrier top.
The propagators $\op{K}_\ell$ and $\op{K}_r$ are thus dominated by states (or paths) localized to the left or right of the dividing surface.
The flux from left to right is
$\op{F} = \frac{\iu}{\hbar}[\op{H},\op{P}_r] = \frac{1}{2m} [\updelta(\op{x}-x_\sigma)\,\op{p} + \op{p}\, \updelta(\op{x}-x_\sigma)]$.  
The multidimensional generalization follows from $\op{P}_r=\theta(\sigma(\op{x}))$, where $\sigma(x)=0$ defines the dividing surface.

In particular, we will take the low-temperature  ($\beta\rightarrow\infty$) limit of the projected flux correlation function.
We will first consider this expression applied to an effective two-level quantum-mechanical system to determine its relation to the tunneling frequency,
before matching the results with a semiclassical approximation to the path-integral formulation of the same quantity.

\subsection{Two-level system}
\label{sec:Cff}

\begin{figure}
    \centering
    \includegraphics[width=0.45\textwidth]{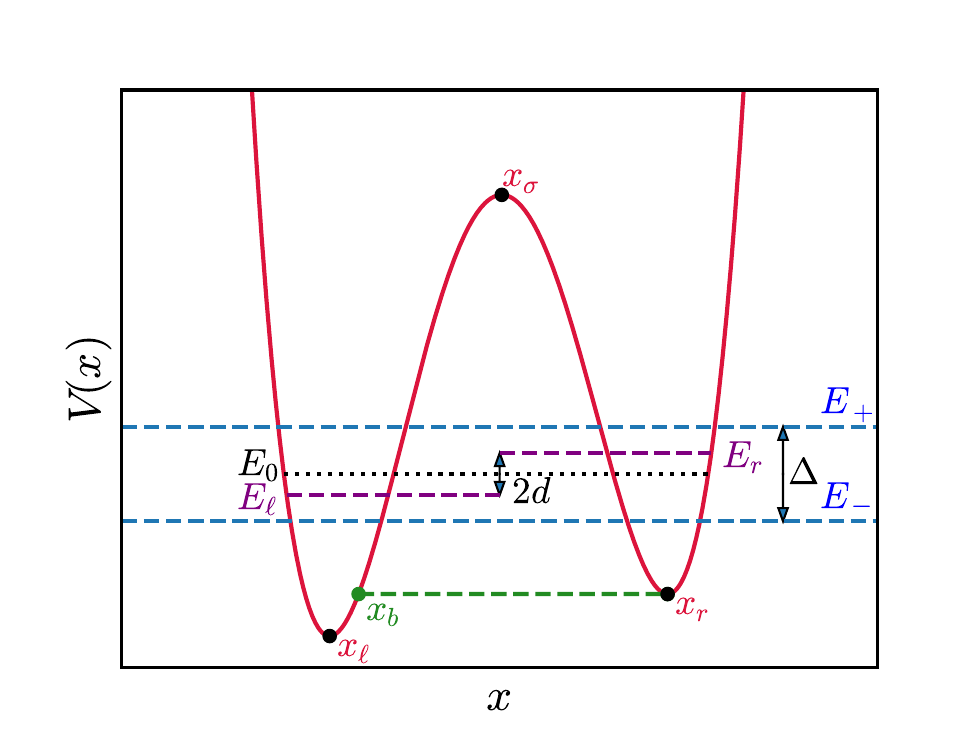} 
    \caption{A double-well potential for which the minima $x_{\ell/r}$, the location of the dividing surface $x_\sigma$, and bounce point $x_b$ are indicated.
    The localized states have energies $E_{\ell/r}$, whereas the eigenstates have energies $E_\pm$. The level splitting, $\Delta$, is determined by the asymmetry, $2d$, in addition to the tunneling frequency.
    }
    \label{fig:doublewell}
\end{figure}

In the low-temperature limit, a double-well potential (depicted in Fig.~\ref{fig:doublewell}) can be described as 
the effective two-level system,
\begin{equation}
    \mathcal{H} =
    \begin{pmatrix}
    E_\ell & -\hbar \Omega \\
    -\hbar \Omega & E_r
    \end{pmatrix}
    =
    \begin{pmatrix}
    E_0-d & -\hbar \Omega \\
    -\hbar \Omega & E_0+d 
    \end{pmatrix} ,
\end{equation}
where we have defined
$E_0 = \thalf(E_\ell + E_r)$ and $d = \thalf(E_r - E_\ell)$
given the energies of the localized states $E_\ell$ and $E_r$.
The $\ell/r$ subscripts refer to objects localized in the left and right wells, 
and $\Omega$ is the tunneling frequency that we will determine later using instanton theory. Given the effective Hamiltonian, the dynamics of this two-level system are simple to describe. \cite{asymtunnel,FeynmanLecturesIII}
In particular, diagonalizing it yields the eigenvalues $E_\pm=E_0\pm\sqrt{d^2+(\hbar\Omega)^2}$, 
so that the level splitting (defined as the difference between these energy levels) is given by $\Delta = E_+ - E_- = 2 \sqrt{d^2+(\hbar \Omega)^2}$; see also Fig.~\ref{fig:doublewell}. The eigenfunctions of this Hamiltonian can be written as $\psi_{-}=(\cos{\tfrac{\phi}{2}}, \sin{\tfrac{\phi}{2}})$ and $\psi_{+}=(-\sin{\tfrac{\phi}{2}}, \cos{\tfrac{\phi}{2}})$, with the mixing angle $\phi=\arctan\frac{\hbar \Omega}{d}$.

We now wish to evaluate $c_\text{ff}(\tau)$ for the two-level system.
For this we use the effective Hamiltonian $\mathcal{H}$ as well as the projection and flux operators
in the localized basis
\begin{subequations}
\begin{align}
    \mathcal{P}_\ell &= \begin{pmatrix} 1 & 0 \\ 0 & 0 \end{pmatrix},
    \qquad
    \mathcal{P}_r = \begin{pmatrix} 0 & 0 \\ 0 & 1 \end{pmatrix}, \label{eq:P2LS}
    \\
    \mathcal{F} &= \frac{\iu}{\hbar} [\mathcal{H}, \mathcal{P}_r]
    =
    \begin{pmatrix} 0 & -\iu\Omega \\
         \iu\Omega & 0 \end{pmatrix}. 
\end{align}
\end{subequations}
These are simply the matrix representations of the operators defined before within the two-level system. 
We have implicitly made the approximation that the basis wavefunctions 
are localized so strongly on the left/right that we can neglect any contribution which leaks onto the other side.
This assumption is only valid if the dividing surface is chosen intelligently (close to the barrier top) and as long as the barrier is high enough.
High barriers are also required for standard instanton theory to work,\cite{tunnel} and so they do not introduce additional limitations.

We can now evaluate the trace in $c_\text{ff}(\tau)$. After some algebra, we find
\begin{equation}
\begin{aligned}
    c_\text{ff}(\tau) &= \Tr \big[ \mathcal{F} \, \eu{-\frac{1}{2\hbar}\tau \mathcal{H}} \mathcal{P}_\ell \, \eu{-\frac{1}{2\hbar}\tau \mathcal{H}}
    \\
    & \hspace{1cm}\, \times \mathcal{F} \, \eu{-\frac{1}{2\hbar}(\beta\hbar-\tau) \mathcal{H}} \mathcal{P}_r \, \eu{-\frac{1}{2\hbar}(\beta\hbar-\tau) \mathcal{H}} \big]
    \\ &\simeq \Omega^2 \, \eu{-\tau E_\ell/\hbar} \, \eu{-(\beta\hbar-\tau) E_r/\hbar},
    \label{eq:cff_twolevel}
\end{aligned}
\end{equation}
where we have retained only the terms to lowest order in $\Omega$.
Note that except in the symmetric case (where $E_\ell=E_r$) the correlation function depends on $\tau$.
As this formulation does not give a preference for any particular choice of $\tau$,
we are free to select the value based on the instanton analysis carried out in Sec.~\ref{sec:1d-instanton}.

Finally, we write the expression in a basis-independent form
\begin{align}     \label{CffPQM}
    c_\text{ff}(\tau) &\simeq \Omega^2 Z_\ell(\tau) Z_r(\beta\hbar-\tau),
\end{align}
which is valid in the limits $\omega_\ell \tau\gg1$ and $\omega_r(\beta\hbar-\tau)\gg1$.
\subsection{Instanton theory}

We have shown in Sec.~\ref{sec:Cff} that the low-temperature limit of the the projected flux correlation function [\eqn{CffPQM}] has a simple connection to the tunneling frequency $\Omega$.
Given this, we shall now apply the instanton formulation to approximate the correlation function, and hence obtain an estimate for $\Omega$.
In contrast to the previous section, we now work in the position representation using the full system Hamiltonian, $\hat{H}$.

\subsubsection{Semiclassical partition functions}
We start out with a short recapitulation of the semiclassical approximation, and in particular obtain an expression for the partition function, which we will use later on in Sec.~\ref{sec:1d-instanton}.

The semiclassical approximation to the imaginary-time propagators $K(x',x'',\tau)=\bra{x''}\exp (-\tau \hat{H}/\hbar) \ket{x'}$ in \mbox{$f$-dimensional} space is \cite{Miller1971density}
\begin{align} \label{eq:kernel-continuous}
    K(x',x'',\tau) \simeq \sqrt\frac{C}{(2\pi\hbar)^f} \, \eu{-S(x',x'',\tau)/\hbar},
\end{align}
where $S(x',x'',\tau)=\int_0^\tau [\half m ||\dot{x}(u)||^2 + V(x(u))] \rmd u$ is the Euclidean action along a classical trajectory which travels from $x(0)=x'$ to $x(\tau)=x''$ in imaginary time $\tau$.
The trajectory is defined as a local minimum of $S$ and if there is more than one solution (e.g., one which bounces on the left and another which bounces on the right), one should sum \eqn{eq:kernel-continuous} over each solution. 
The prefactor accounts for fluctuations around the trajectory and is defined as the $f\times f$ determinant
$C=\left|-\pders{S}{x'}{x''}\right|$.
The projection operator $\op{P}_{\ell/r}$, which appears in the definition of the propagators $\op{K}_{\ell/r}$, is easily accounted for within the semiclassical theory;
one simply discards the trajectories which are on the wrong side of the dividing surface at their half-way point.
The projected propagators are thus simpler to handle than the standard unprojected versions, which typically depend on two trajectories. \cite{QInst}
The energy of an imaginary-time classical trajectory is defined by $E = \pder{S}{\tau}$, which evaluates to a conserved quantity, $E=-\half m ||\dot{x}(u)||^2 + V(x(u))$ along the trajectory.

The full power of the semiclassical propagator will become apparent later.
First, as a useful exercise, we use it to formulate an approximation to the partition functions of the individual wells. 
For notational convenience, we define
$\tau_\ell\equiv\tau$ and $\tau_r\equiv\beta\hbar-\tau$.
The partition function of the left well is approximated  by \cite{InstReview}
\begin{equation} \label{eq:well-partition}
    Z_\ell(\tau_\ell) = \int K_\ell(x,x,\tau_\ell) \, \rmd x \simeq K_\ell^{(0)}(\tau_\ell) \,\Xi_\ell^{(0)}(\tau_\ell) ,
\end{equation}
where the relevant classical path is collapsed at the bottom of the well (i.e.\ $x(u)=x_{\ell}$), such that
the semiclassical propagator is given by  
\begin{align} \label{eq:sc-trap}
    &K_\ell^{(0)}(\tau_\ell)=\sqrt{\frac{C_\ell^{(0)}}{(2\pi\hbar)^f}} \,
    \eu{-S^{(0)}_\ell / \hbar}, 
\end{align}
and the term resulting from the steepest-descent integration over the end points is given by 
\begin{align} \label{eq:trap-xi}
    &\Xi^{(0)}_\ell(\tau_\ell) = (2\pi\hbar)^{f/2} 
    \left|\pders{S_\ell^{(0)}}{x}{x}\right|^{-1/2}.
\end{align}
Here, the $(0)$ superscripts indicate that the quantity corresponds to that of a collapsed trajectory, e.g.\ $S_\ell^{(0)}=S(x_\ell, x_\ell, \tau_\ell)=\tau_\ell V(x_\ell)$, and
$\pders{S_\ell^{(0)}}{x}{x} = \pders{S_\ell}{x'}{x'} + \pders{S_\ell}{x''}{x''} + \pders{S_\ell}{x'}{x''} + \pders{S_\ell}{x''}{x'}$.

These formulas reproduce the quantum-mechanical partition functions within the harmonic approximation. \cite{InstReview}
For instance, in one dimension,
\begin{align} \label{eq:z-trap-exact}
    &Z_\ell \simeq \left(2\sinh{\thalf\tau_\ell\omega_\ell}\right)^{-1} \eu{-\tau_\ell V_\ell / \hbar}, 
\end{align}
where
$V_\ell=V(x_\ell)$ and
$\omega_\ell$ is the harmonic vibrational frequency of the left well.
From this result, we can assign the semiclassical zero-point energy as ${E}_{\ell} = V_{\ell} + \half \hbar \omega_{\ell}$.
Finally, it should be noted that Eqs.~(\ref{eq:well-partition}--\ref{eq:z-trap-exact}) can also be applied to the right way after making the trivial substitutions. 


\subsubsection{Semiclassical derivation} \label{sec:1d-instanton}
We now move on to the derivation of the semiclassical approximation to the correlation function.
For the sake of clarity, we start with the one-dimensional case; we generalize the theory to multidimensional problems in Sec.~\ref{subsec:rp-inst}.

By expanding the trace in \eqn{eq:cff_definition} in a basis of position states and evaluating the momenta as described in Ref.~\onlinecite{InstReview},
we obtain 
\begin{align}
    c_\text{ff}({\tau})
    &\simeq \int \text{d}x' \int \text{d}x'' \, |\dot{x}'| |\dot{x}''| \updelta(x'-x_\sigma) \updelta(x''-x_\sigma) \nonumber \\ 
    & \hspace{2cm} \times K_\ell(x',x'',\tau_\ell) K_r(x'',x',\tau_r) \nonumber \\
    &\simeq  |\dot{x}_\sigma|^2 K_\ell(x_\sigma, x_\sigma, \tau_\ell) K_r(x_\sigma, x_\sigma, \tau_r) .
     \label{eq:cff-inst}
\end{align}
Here we note that we have imaginary-time propagators corresponding to two trajectories starting and ending at $x_\sigma$; one which bounces
on the left-hand side of the dividing surface and another which bounces on the right-hand side.
We have chosen to evaluate the correlation function at the unique value of $\tau$ for which the energies of the two trajectories match, $E_\ell=E_r$, so that together they form a periodic orbit. This simplifies the implementation later on, as it will allow us to use standard ring-polymer instanton optimization routines.\cite{InstReview} Additionally, it will enable the connection to rate theory made in Sec.~\ref{subsec:discuss-rate}.

\begin{figure}[h!] 
    \centering
    {\includegraphics[width=0.9\linewidth]{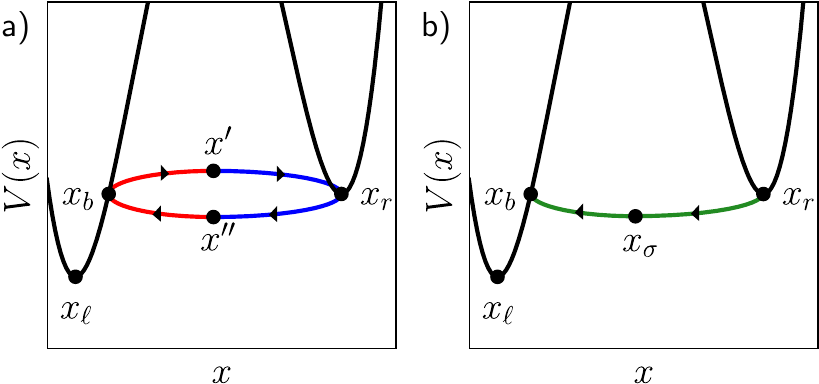}}  
    \caption{Trajectories of interest in the 1D asymmetric double well. a) Representation of the left (red) and right (blue) bouncing trajectories defined in Eq.~\eqref{eq:cff-inst}; b) Representation of the trajectory associated with $K_\mathrm{pin}$ defined in Eq.~\eqref{eq:Kpin}.}
    \label{fig:trajectories}
\end{figure}

The two bouncing trajectories from \eqn{eq:cff-inst} are illustrated in Fig.~\ref{fig:trajectories}a.
In the low-temperature limit, the blue path starts at $x'=x_\sigma$ and reaches the bottom of the right well, $x_r$, where it can spend an arbitrarily long amount of imaginary time before returning to the dividing surface.
Because the left-bouncing path has the same energy,
the red path travels from $x''=x_\sigma$ towards the left (deeper) well, but does not reach the bottom.
Instead, due to energy conservation, 
it bounces off the turning point $x_b$, at which $V(x_b)=V(x_r)$, 
and then returns to $x'=x_\sigma$ in a finite amount of imaginary time $\tau$.

To calculate $c_\text{ff}(\tau)$ numerically, it will be helpful to reformulate \eqn{eq:cff-inst}. To do this, we start by realizing that as a result of the projection on the left/right localized states, 
the left- and right-bouncing paths can be split once each at their half-way points $x_b$ and $x_r$ such that
\begin{subequations}
\label{eq:kernels-pm}
\begin{align} 
    K_\ell &= K(x_\sigma, x_b, \thalf\tau_\ell) \Xi_{\sigma b} K(x_b, x_\sigma, \thalf\tau_\ell)
    \\
    K_r &= K(x_\sigma, x_r, \thalf\tau_r) \Xi_{\sigma r} K(x_r, x_\sigma, \thalf\tau_r) .
\end{align}
\end{subequations}
The extra factors account for fluctuations of the point at which the path was split and are given by \cite{GutzwillerBook,InstReview}
\begin{subequations}
\begin{align}
    &\Xi_{\sigma b} = \left({2\pi\hbar}\right)^{1/2} \left|2\pders{S_{\sigma b}}{x_b}{x_b}\right|^{-1/2}  \\
    &\Xi_{\sigma r} = \left({2\pi\hbar}\right)^{1/2} \left|2\pders{S_{\sigma r}}{x_r}{x_r}\right|^{-1/2} ,
\end{align}
\end{subequations}
where $S_{\sigma b} = S(x_\sigma, x_b,\half\tau_\ell)$ is the action associated with the trajectory traveling from $x_\sigma$ to $x_b$, and similarly, $S_{\sigma r} = S(x_\sigma,x_r,\half\tau_r)$ is the action for the trajectory traveling from $x_\sigma$ to $x_r$.

As $K(x_\sigma, x_b, \frac{1}{2}\tau_\ell) = K(x_b, x_\sigma, \frac{1}{2}\tau_\ell)$ (and similarly for the two trajectories on the right), we can now write 
\begin{align}
    c_\text{ff} &\simeq   |\dot{x}_\sigma|^2  \Xi_{\sigma b} \Xi_{\sigma r} K_\text{pin}^2,  
\end{align} 
where $K_\text{pin}$ is defined as (for $f=1$)
\begin{align}
    \label{eq:Kpin}
    K_\text{pin}
    &=K(x_b, x_\sigma, \thalf\tau_\ell) K(x_\sigma, x_r, \thalf\tau_r) \nonumber \\
    &= \sqrt{\frac{C_{\sigma b}}{2\pi\hbar}} \sqrt{\frac{C_{\sigma r}}{2\pi\hbar}} \, \eu{-S_\text{inst} / \hbar}.
\end{align}
To obtain $K_\text{pin}$, one would now only need one trajectory traveling from $x_b$ to $x_r$ in time $\tau_\text{inst}=\half(\tau_\ell+\tau_r)=\half\beta\hbar$ with total action $S_\text{inst}=S(x_b,x_\sigma,\half\tau_\ell)+S(x_\sigma,x_r,\half\tau_r)$; see Fig.~\ref{fig:trajectories}b. 
It is this path which we shall call the `instanton' trajectory.
Note that once the trajectory is found, it is easier to work with the two parts separately, as implied by Eq.~\eqref{eq:Kpin}.
The prefactor of $K_\text{pin}$ is different from the semiclassical propagator from $x_b$ to $x_r$; it does not allow fluctuations at the pinned point $x_\sigma$, which thus obeys the delta-function conditions and avoids the complication that the direct path has a zero mode in the symmetric case.

Now that we have obtained a workable expression for $c_\text{ff}$, we combine it with the expression derived from the effective two-level system [\eqn{CffPQM}] to write $\Omega$ as
\begin{align}\label{eq:Omega_asym}
    \Omega &\simeq |\dot{x}_\sigma| K_\text{pin} \sqrt\frac{\Xi_{\sigma b} \Xi_{\sigma r}}{Z_\ell Z_r}.
\end{align}
Using the semiclassical formulation of the partition function introduced in the previous section, we obtain the final instanton expression for the tunneling frequency:
\begin{align}
    \label{eq:Omega_asym_expanded}
    \Omega \simeq \Gamma(\tau) |\dot{x}_\sigma| \sqrt{
    \frac{C_{\sigma b} C_{\sigma r}}{2\pi\hbar \left(C^{(0)}_\ell C_r^{(0)}\right)^\half  }
    } \, \eu{-(S_\text{inst} - \half\tau_\ell V_\ell - \half\tau_r V_r) / \hbar},
\end{align}
where
\begin{align}
    \Gamma(\tau) = \sqrt\frac{\Xi_{\sigma b} \Xi_{\sigma r}}{\Xi_\ell^{(0)} \Xi_r^{(0)}}.
\end{align}

Finally, it is possible to show that the new generalized formula reduces to our previous theories in special cases. 
In particular, \eqn{eq:Omega_asym_expanded} reduces to the result obtained in Ref.~\citenum{asymtunnel} for a case where only the widths of the wells are asymmetric and $V_\ell = V_r$ (but not necessarily $E_\ell = E_r$). 
In this case, the instanton trajectory obeys $x_b = x_\ell$. 
By making a comparison to Eq.~(90) of Ref.~\citenum{InstReview}, we recognize that in this case,
\begin{align}
    K_1'(\tau_\text{inst}) = |\dot{x}_\sigma|  K_\text{pin}(\tau_\text{inst}). 
\end{align}
We can then rewrite Eq.~(\ref{eq:Omega_asym_expanded}) 
as 
\begin{align}
    \Omega \simeq \Gamma(\tau) \frac{K_1'(\tau_\text{inst})}{\sqrt{K_\ell(\tau_\ell) K_r(\tau_r)}}.
\end{align}
In the limit of long $\tau$, $\Gamma \rightarrow 1$ because both $\Xi_{\sigma r} \rightarrow \Xi_r^{(0)}$ and $\Xi_{\sigma b}=\Xi_{\sigma\ell} \rightarrow \Xi_\ell^{(0)}$. Thus our current formulation reduces to that in Ref.~\citenum{asymtunnel}. By extension, it therefore also reduces to the standard instanton result for a fully symmetric system where $E_\ell=E_r$.\cite{InstReview}  

\subsubsection{Ring-polymer instanton formulation} \label{subsec:rp-inst}
In the previous section, we have derived a instanton expression for the tunneling splitting in one dimension. We now generalize this approach to a multidimensional system. This we do in the framework of discretized path integrals, giving us a ring-polymer instanton formulation\cite{InstReview, asymtunnel, tunnel} which facilitates a straightforward numerical implementation. 

From here on, we consider an $f$-dimensional system which has been mass-weighted such that all degrees of freedom have the same mass, $m$. 
We use a coordinate system $\mathbf{x}=(q,\mathbf{Q})$ defined 
by an orthogonal transform such that the
instanton trajectory is normal to the dividing surface $\sigma(q)=q-q_\sigma$ as it passes through.
The projection and flux operators are defined as before, but in this coordinate system, they depend on the $q$ coordinates only.
The multidimensional expression for the flux correlation function thus reads
\begin{equation} \label{eq:multidim-cff}
\begin{aligned}
    c_\text{ff}({\tau}) = 
    \int \text{d}\mathbf{x}' &\int \text{d}\mathbf{x}''\;
    |\dot{q}'| |\dot{q}''| \delta{(q'-q_\sigma)} \delta{(q''-q_\sigma)} \\
    &\times K_\ell(\mathbf{x}',\mathbf{x}'',\tau_\ell) K_r(\mathbf{x}',\mathbf{x}'',\tau_r).
    \end{aligned}
\end{equation}

We now use the ring-polymer discretization scheme to write the propagator on the left as
\begin{equation} \label{eq:rp-kernel}
\begin{aligned}
    K_\ell(\mathbf{x}_0, \mathbf{x}_{N_\ell}, \tau_\ell) = & \left(\frac{m}{2\pi\epsilon_\ell\hbar}\right)^{N_\ell f/2} \\ &\hspace{-0.15cm} \times \int \text{d}\mathbf{x}_1 \cdots \hspace{-0.09cm}\int \text{d}\mathbf{x}_{N_\ell-1} \, \eu{-S^{(\ell)}_N/ \hbar},
    \end{aligned}
\end{equation}
where the trajectory 
is discretized into $N_\ell$ segments of imaginary time $\epsilon_\ell = \tau_\ell/N_\ell$, 
with each bead $\mathbf{x}_i$ corresponding to a replica of the system,
and $S^{(\ell)}_N$ is the action of the discretized path:
\begin{equation}
    \begin{aligned}
            S^{(\ell)}_N =& \sum_{i=1}^{N_\ell} \frac{m}{2\epsilon_\ell} ||\mathbf{x}_{i} - \mathbf{x}_{i-1}||^2 \\
    &+ 
    \epsilon_\ell \left[
    \half V(\mathbf{x}_0) + \sum_{i=1}^{N_\ell-1} V(\mathbf{x}_i) + \half V(\mathbf{x}_{N_\ell})
    \right].
    \end{aligned}
\end{equation}
The propagator on the right is written similarly.

\begin{figure}
    \centering \includegraphics[width=0.7\linewidth]{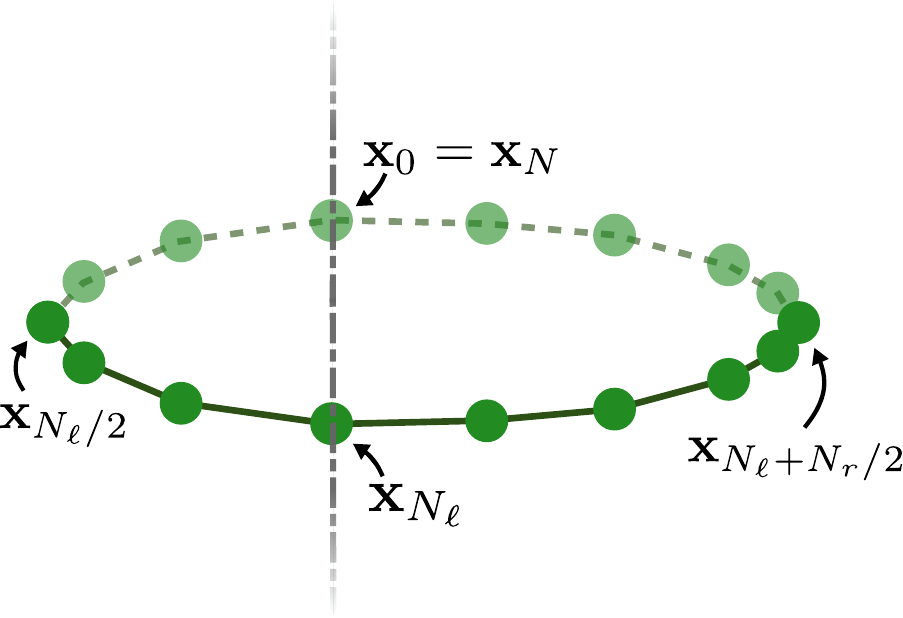}
    \caption{Schematic illustration of the bead indexing convention of the ring-polymer, as described in the main text. The dividing surface is indicated by the vertical line. In practice, we only need to optimize half a ring polymer, indicated by the dark green dots (beads); the positions of the semi-transparent beads follows from the opaque ones.  }
    \label{fig:RPindices}
\end{figure}

We can now use Eq.~\eqref{eq:rp-kernel} to rewrite the flux correlation function in Eq.~\eqref{eq:multidim-cff} as
\begin{align} \label{eq: cff-rp}
    c_\text{ff}({\tau}) &= 
    \left(
    \frac{m}{2\pi\epsilon \hbar}
    \right)^{{Nf}/{2}} \int \text{d}\bm{x} \;
    |\dot{q}'| |\dot{q}''| \delta{(q'-q_\sigma)} \nonumber \\ &\hspace{3.5cm} \times \delta{(q''-q_\sigma)} \; \eu{-S_N(\bm{x}) / \hbar} \nonumber \\
    &\simeq \left(\frac{m}{2\pi\epsilon \hbar}\right)
    \frac{|\dot{q}_\sigma|^2\,\eu{-S_N(\tilde{\bm{x}}) / \hbar}}{\det(\bm{J}_\text{pin})^{\half}}.
\end{align}
Here, the periodic orbit is discretized into $N = N_\ell + N_r$ segments, where $N_\ell$ and $N_r$ correspond to the number of segments of the left- and right-bouncing trajectories respectively, and are chosen such that $\tau_\ell/N_\ell=\tau_r/N_r=\epsilon$; the total action is given by $S_N = S^{(\ell)}_N + S^{(r)}_N$.
The trajectory forms a full ring $\bm{x}=(\mathbf{x}_1,\dots,\mathbf{x}_N)$ such that $\mathbf{x}_0 \equiv
\mathbf{x}_N$, where $\mathbf{x}_0\equiv\mathbf{x}'$ and $\mathbf{x}_{N_\ell}\equiv\mathbf{x}''$; see also Fig.~\ref{fig:RPindices}.
The second line of Eq.~\eqref{eq: cff-rp} is the result of steepest-descent integration around
$\tilde{\bm{x}}$, which corresponds to the optimized ring-polymer instanton configuration.
The matrix $\bm{J}_\text{pin}$ is a result of the steepest-descent integration and is equal to $\epsilon/m$ times the $Nf\times Nf$ matrix of second derivatives of the action, $\frac{\epsilon}{m}\nabla^2 S_N (\tilde{\bm{x}})$, 
with the two rows and columns corresponding to $q', q''$ (or using bead indices, $q_0$ and $q_{N_\ell}$) removed.

Following this, we employ Eq.~\eqref{eq:rp-kernel} to write the partition functions defined in Eq.~\eqref{eq:well-partition} in their discretized form
\begin{subequations}
\begin{align}
    Z_\ell(\tau_\ell) \simeq \frac{\eu{-\tau_\ell V_\ell / \hbar}}{\det(\bm{J}_\ell)^{1/2}}  \\
    Z_r(\tau_r) \simeq \frac{\eu{-\tau_r V_r / \hbar}}{\det(\bm{J}_r)^{1/2}},
\end{align}
\end{subequations}
where $\bm{J}_\ell = \frac{\epsilon}{m} \nabla^2 S_N(\bm{x}_\ell)$ and $\bm{J}_r = \frac{\epsilon}{m} \nabla^2 S_N(\bm{x}_r)$.
Here, $\bm{x}_\ell$ and $\bm{x}_r$ correspond to ring polymers which are collapsed on the left and right well, respectively.

Finally, we use Eq.~\eqref{CffPQM}, to obtain $\Omega$ as
\begin{align}\label{eq:omega-rp}
    \Omega &\simeq  \sqrt{\frac{c_\text{ff}}{Z_\ell Z_r}} 
    =\frac{1}{\Phi_\text{pin}} \frac{|\dot{q}_\sigma|}{\sqrt{2\pi\hbar}} \, \eu{-(S_\text{inst} - \half \tau_\ell V_\ell - \half \tau_r V_r)/\hbar },
\end{align}
where $S_\text{inst}=\half S_N(\tilde{\bm{x}})$, and 
\begin{align}
    \Phi_\text{pin} = \sqrt{\frac{\epsilon}{m}} \left[\frac{\det{(\bm{J}_\text{pin})}}{\det{(\bm{J}_\ell)} \det{(\bm{J}_r)}}\right]^{1/4}.
\end{align} 

We thus form a full ring-polymer trajectory by concatenating the single trajectory with its reversed version in order to obtain the fluctuation factor $\Phi_\text{pin}$. The dividing surface is chosen after instanton optimization at the bead with the largest potential.



While the starting premise of Eq.~\eqref{eq:omega-rp} is different from the standard instanton result for symmetric systems,
it can be shown that we recover this standard result in the symmetric case.
To relate the projected flux correlation approach presented in this work to the partition-function based approach of the original instanton formulation,
we turn to the work by Althorpe. \cite{Althorpe2011ImF}
By using Eqs.~(24--27) of Ref.~\citenum{Althorpe2011ImF},
we obtain an alternative formula for $\Phi_\text{pin}$ valid for symmetric systems as
\begin{align}
    \Phi_\text{pin} = \frac{\epsilon|\dot{q}_\sigma|}{\sqrt{S_\text{inst}}}  \left[
    \frac{\det''(\bm{J})}{\det(\bm{J}_0)}
    \right]^\frac{1}{4},
\end{align}
where the double prime symbol indicates the removal of two zero modes in the calculation of the determinant of the two-kink ring polymer.\cite{tunnel}
Inserting this form of $\Phi_\text{pin}$ into Eq.~\eqref{eq:omega-rp} allows us to recover the standard instanton result for symmetric systems. 




\section{Results and Applications} \label{sec:method}


In this section we first highlight some important points concerning the computational algorithm used to implement the ring-polymer formulation of projected-flux instanton theory.
We then apply it to two model systems to compare with benchmark results
and also to an asymmetric biomolecule, \textalpha-fenchol, \cite{fenchol} in full dimensionality.

As in other instanton methods,\cite{RPInst,Andersson2009Hmethane,Rommel2011locating}
it is not necessary to optimize the full periodic orbit using a full ring polymer because the path folds back on itself.  Therefore, we need only consider 
one passage of the barrier (as indicated in Fig.~\ref{fig:RPindices}) traveling
in half the imaginary time $\tau_\text{inst} = \beta \hbar / 2$. 
This path is discretized into $N_\text{inst}$ beads and then optimized such that it is a \emph{saddle point} of the action. 
This is a key difference from the standard instanton approach for tunneling splittings in symmetric systems or even asymmetric isotopically substituted molecules, 
for which the instanton is a minimum of the action. \cite{tunnel, asymtunnel, InstReview}
Our problem is therefore more similar to that of ring-polymer instanton rate theory, \cite{InstReview,RPInst,Andersson2009Hmethane} 
and we accordingly exchanged the usual {l-BFGS} algorithm for Newton--Raphson and other eigenvector-following algorithms to obtain the optimized instanton trajectory.

From this pathway, we can form a full ring polymer by concatenating the single trajectory with its reversed version, from which we can obtain the fluctuation factor $\Phi_\text{pin}$. The dividing surface is chosen after instanton optimization at the bead with the largest potential.

\subsection{A 1D asymmetric double well}
In order to benchmark our new theory, we modify the usual symmetric quartic double-well potential \cite{ABCofInstantons, Benderskii, Polyakov1977instanton} by introducing a slight asymmetry to the system such that $V_\ell < V_r$. 
We therefore define the one-dimensional asymmetric double-well potential as
\begin{align} \label{eq:1d-dw}
    V(x)=V_0\Bigg[\bigg(\frac{x^2}{x_0^2}-1\bigg)^2 + a\frac{x}{x_0}\Bigg]
\end{align}
where $a\ge0$ is a parameter which determines the asymmetry of the system.
The barrier height is $V_0$, the length scale is chosen to be $x_0=5\sqrt{V_0}$
and we employ reduced units such that $m=1$ and $\hbar=1$.
These parameters are chosen similarly to our previous work. \cite{tunnel,asymtunnel} The harmonic frequencies are given by $\omega_{\ell/r} = \sqrt{\nabla^2 V(x_{\ell/r})/m}$ 
and for small $a$, $d \approx 2a + \hbar(\omega_r - \omega_\ell)/4$. 

\begin{table*}[t]
\centering
\setlength{\tabcolsep}{12pt} 
\begin{tabular}{lllllll}
\hline\hline
$V_0$ & $a$    & $d$     & $\Omega_\text{inst}$ & $\Delta_\text{inst}$ & $\Delta_\text{exact}$ & Error {[}\%{]} \\ \hline
0.5   & 1.($-$1) & 4.5($-$2) & 1.71($-$2)             & 9.64($-$2)             & 8.80($-$2)              & 9.5           \\
1     & 1.($-$3) & 9.5($-$4) & 1.94($-$4)             & 1.92($-$3)             & 1.91($-$3)              & $< 1$         \\
2     & 1.($-$7) & 2.0($-$7) & 2.20($-$8)             & 3.92($-$7)             & 3.90($-$7)              & $< 1$         \\ \hline\hline
0.5   & 1.($-$2) & 4.5($-$3) & 1.55($-$2)             & 3.23($-$2)             & 2.41($-$2)              & 34            \\
1     & 1.($-$4) & 9.5($-$5) & 1.93($-$4)             & 4.31($-$4)             & 3.91($-$4)              & 10            \\
2     & 1.($-$8) & 2.0($-$8) & 2.20($-$8)             & 5.88($-$8)             & 5.69($-$8)              & 3.3           \\ \hline\hline
0.5   & 1.($-$3) & 4.5($-$4) & 1.53($-$2)             & 3.06($-$2)             & 2.25($-$2)              & 36            \\
1     & 1.($-$5) & 9.5($-$6) & 1.94($-$4)             & 3.87($-$4)             & 3.43($-$4)              & 13            \\
2     & 1.($-$9) & 2.0($-$9) & 2.20($-$8)             & 4.42($-$8)             & 4.17($-$8)              & 6             \\ \hline\hline
\end{tabular}
\caption{
Level splittings obtained from instanton theory $\Delta_\text{inst} = 2\sqrt{d^2 + (\hbar\Omega)^2}$ as well as from exact quantum mechanics $\Delta_\text{exact}$ for the one-dimensional asymmetric double well [Eq.~\eqref{eq:1d-dw}].
The results are presented in three parameter regimes for three different barrier heights $V_0$; the top section represents the high asymmetry regime wherein the parameter $a$ is selected such that $d \gg \hbar\Omega$, the middle section with $d \approx \hbar\Omega$ and the bottom section with $d \ll \hbar\Omega$.
Powers of 10 are given in parentheses.}
\label{tab:1d-results}
\end{table*}

Table \ref{tab:1d-results} presents the results of our new theory and compares them with benchmark quantum-mechanical results obtained from a numerical solution of the Schr\"odinger equation via the discrete variable representation (DVR). \cite{Light1985DVR}
The instanton results were obtained with $N_\text{inst}=2048$ and $\tau_\text{inst}=150$ to ensure full convergence.
We vary the value of $a$ to 
study three regimes:
low ($d \ll \hbar\Omega$), medium ($d \approx \hbar\Omega$) and high asymmetry ($d \gg \hbar\Omega$). 
In the low asymmetry case, the level splittings predicted by our new theory essentially match those of the symmetric case.\cite{tunnel}
The same behavior is mirrored in the quantum results and occurs because the result is dominated by $\hbar\Omega$, 
with only tiny contributions from $d$.
On the other hand, for the highest asymmetry case, 
the level splitting is almost completely dominated by $d$.
In the case where $d \approx \hbar\Omega$, both $d$ and $\hbar\Omega$ contribute significantly to $\Delta$.
Additionally, we note that, for the most part $\hbar\Omega$ remains nearly unchanged with varying asymmetry.
These results are not unexpected; in fact these were the very same trends that we have observed in Ref.~\citenum{asymtunnel}.
However, for $V_0=0.5$, the behavior is not trivial and $\hbar\Omega$ does change for the highest asymmetry case. 
Note that 
one could not have just employed the instanton theories presented in Ref.~\citenum{asymtunnel} or Ref.~\citenum{InstReview} as they are simply not applicable to even weakly asymmetric systems. 
Finally, we remark that in each case, our results compare favorably with those obtained by exact quantum mechanics. 
In particular, the accuracy of the result increases with increasing barrier height, which is where the semiclassical approximation is valid.\cite{Kleinert} 




\subsection{A 2D asymmetric double well}\label{subsec:2d}

\begin{figure}[h!]
    \centering
    \includegraphics[width=\linewidth, keepaspectratio]{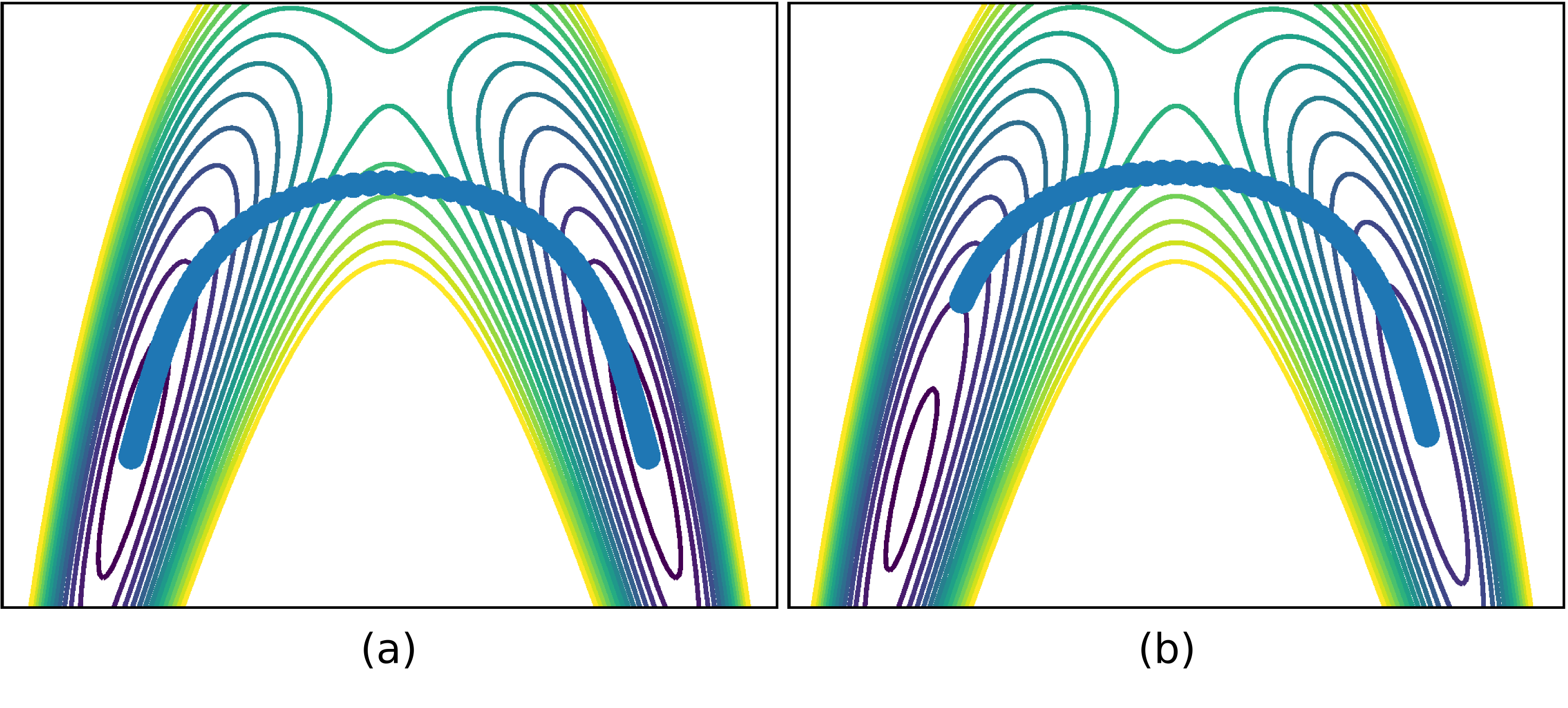} 
    \caption{
    Plots of the 2D PES defined in Eq.~\eqref{eq:2d-dw}, with (a) being fully symmetric (i.e. with $a=0$), 
    and (b) being a highly exaggerated asymmetric system, with $a=0.01$. 
    The instanton trajectory is also shown for each case, with its beads represented as blue circles. Note that the ring polymer folds back on itself.
    }
    \label{fig:2d-pes} %
\end{figure}

In this work, we modify the standard two-dimensional symmetric mode-coupling potential\cite{Benderskii} by introducing a cubic term $ax^3$ such that the wells have asymmetric depths ($V_\ell < V_r$).
This potential is
\begin{align} \label{eq:2d-dw}
    V(x,y) = \frac{1}{8}(x-1)^2(x+1)^2 + \frac{\omega_y^2}{2}\left[y+c(x^2-1)\right]^2 +
    ax^3.
\end{align} 
Here, we set $c=1, m=1, \hbar = 0.04$ and $\omega_y=0.8$, which are the parameters chosen by Mil'nikov and Nakamura. \cite{NakamuraBook}
We then vary $a$ to generate systems with different asymmetric characters.
The fully symmetric PES is presented in Fig.~\ref{fig:2d-pes} along with its highly exaggerated asymmetric counterpart, 
wherein $a$ was chosen to be large enough to clearly show the asymmetry in the PES.

Fig.~\ref{fig:2d-pes} also shows the instanton trajectories obtained by the ring-polymer saddle-point search.
For the asymmetric system, the instanton trajectory starts from the right minimum,
crosses the barrier, reaches a bounce point on the left but never reaches the bottom of the left well.

We present the level splittings calculated for varying degrees of asymmetry in Table \ref{tab:2d}. 
The instanton results were converged with $N_\text{inst}=1024$ and $\tau_\text{inst}=120$.
In all cases, agreement between instanton and quantum-mechanical results is excellent,
with all errors falling below $10 \%$.
The trends in the 2D system effectively mirror those of the 1D example.
For the low asymmetry case, the level splitting is almost identical to that of the fully symmetric case ($a=d=0$).
For the medium asymmetry case, both asymmetry and tunneling make contributions to the level splitting.
For the highest asymmetry case,
the level splitting is almost completely dominated by the asymmetry such that $\Delta \approx 2d$; in this case the results are very accurate because they are not affected by the semiclassical instanton approximations. The small error comes from the harmonic approximation in the wells, which appears to be excellent for this simple model.

\begin{table*}[t]
\centering
\setlength{\tabcolsep}{12pt} 
\begin{tabular}{llllll}
\hline \hline
$a$ & $d$ & $\hbar\Omega$ & $\Delta_\text{inst}$ & $\Delta_\text{exact}$ & Error {[}\%{]} \\ 
\hline
0        & 0        & 2.49($-$9) & 4.99($-$9) & 4.55($-$9)  & 9.7  \\
2.5($-$10) & 2.3($-$10) & 2.50($-$9) & 5.01($-$9) & 4.57($-$9) &  9.6\\
2.5($-$9)  & 2.3($-$9)  & 2.50($-$9) & 6.77($-$9) & 6.43($-$9) &  5.3\\
2.5($-$8)  & 2.3($-$8)  & 2.50($-$9) & 4.60($-$8) & 4.57($-$8) &  $<$ 1\\
\hline \hline
\end{tabular}
\caption{Tunneling splittings $\Omega$ and level splittings $\Delta_\text{inst} = 2\sqrt{d^2 + (\hbar\Omega)^2}$ from instanton theory,  as well as from exact quantum mechanics $\Delta_\text{exact}$ via DVR for the 2D system defined by \eqn{eq:2d-dw}.
The parameter $a$ is varied to achieve different levels of asymmetry, as measured by the relative magnitudes of $d=\thalf\left({E}_r - {E}_\ell\right)$,
and the tunneling frequency $\Omega$,
calculated with ring-polymer instanton theory.
Powers of 10 are given in parentheses.} 
\label{tab:2d}
\end{table*}

\subsection{\textalpha-fenchol} \label{subsec:fenchol}
\textalpha-fenchol is a biomolecule found in fennel,  
from which it takes its name.
Moreover, it gives basil its characteristic scent and is extensively used in perfumery.
Rotational microwave jet spectroscopy studies by Medel et al. \cite{fenchol} 
observed a spectral splitting, which they assign to the tunneling between two nearly degenerate conformers $g-$ and $g+$ (see Fig.~2 of Ref.~\citenum{fenchol}).
Additionally, they have conducted a simple theoretical investigation wherein they used a 1D model in combination with dispersion-corrected B3LYP and CCSD(T) calculations.
This approach predicted that the tunneling and asymmetric contributions to the level splitting were of the same order of magnitude.
This is therefore a perfect multidimensional system on which to demonstrate our new asymmetric generalization of instanton theory.

To reduce the computational costs associated with on-the-fly calculations,
we employed the machine-learning method of Gaussian process regression (GPR). \cite{GPRbook}
We have shown in previous work \cite{GPR,Muonium,newGPR} that the GPR-aided instanton approach can reduce the number of electronic-structure calculations required by at least an order of magnitude without significantly impacting the result of a thermal rate calculation.
The speed-up is expected to be even larger for the present approach, as
tunneling-splitting calculations are evaluated in the low-temperature limit and thus use even longer trajectories than in instanton rate theory. This means that the instanton needs to be represented by a far larger number of beads, which increases the number of potential, gradient and Hessian computations required.
In combination with the large number of degrees of freedom presented by \textalpha-fenchol, this would present a high computational cost for conventional on-the-fly instanton theory.  However, 
in this work we demonstrate that the GPR-aided instanton approach allows us to evaluate the level splitting with only moderate resources.

The GPR-based PES was built from an initial training dataset which consists of a few potentials and gradients obtained by B3LYP-D3BJ/6-311++G(d,p) calculations at points based on an initial guess configuration of the instanton.
We additionally included the potentials, gradients and Hessians of the $g+$ and $g-$ minima as well as the transition state.
All electronic-structure calculations were performed using ORCA. \cite{Neese2012orca}
As in previous work, the GPR-based PES is refined through the addition of more potentials, gradients and Hessians until a converged instanton trajectory and its associated value of $\hbar\Omega$ is obtained.
Our converged value for $\hbar\Omega$ at the B3LYP-D3BJ/6-311++G(d,p) level of theory was obtained using a GPR-based PES composed of 50 potentials, gradients and 8 Hessians. 
The instanton was discretized with $N_\text{inst}=2048$ beads, 
and optimized at an `effective temperature' of $T_\text{inst} = \hbar/(k_\mathrm{B} \tau_\text{inst}) = 40 \: \mathrm{K}$. 
Here, the advantage of the GPR-aided instanton approach becomes evident: we only required 8 Hessians to obtain the converged values for $\hbar\Omega$ as opposed to 2048 Hessians that we would have calculated using an on-the-fly approach.
To increase the accuracy of the value obtained for $\hbar\Omega$, $d$, and hence the level splitting $\Delta$,
we further employ a dual-level approach, \cite{Milnikov2003,Meisner2018dual,nitrene,tropolone}
wherein we correct the potentials along the instanton obtained from the DFT-based GPR PES with DLPNO-CCSD(T)/\text{aug-cc-pVQZ}.\cite{Neese2012orca}

In addition to the GPR-aided instanton approach,
we also need to employ to the divide-and-conquer scheme for the calculation of the fluctuation factors.\cite{DaC}
This is because the large number of degrees of freedom of \textalpha-fenchol cause the diagonalization of $\bm{J}_\text{pin}$ to be far too computationally expensive,
even with banded-matrix algorithms. On the other hand, the divide-and-conquer scheme avoids diagonalization and could be evaluated reasonably quickly.
We find that for the fully converged instanton at $T_\text{inst}=40$ K discretized with $N_\text{inst}=2048$ beads, 
splitting the instanton trajectory 16 times was sufficient to achieve numerical stability in $\Omega$.


\begin{figure}[h!]
    \centering
    \includegraphics[width=0.7\linewidth, keepaspectratio]{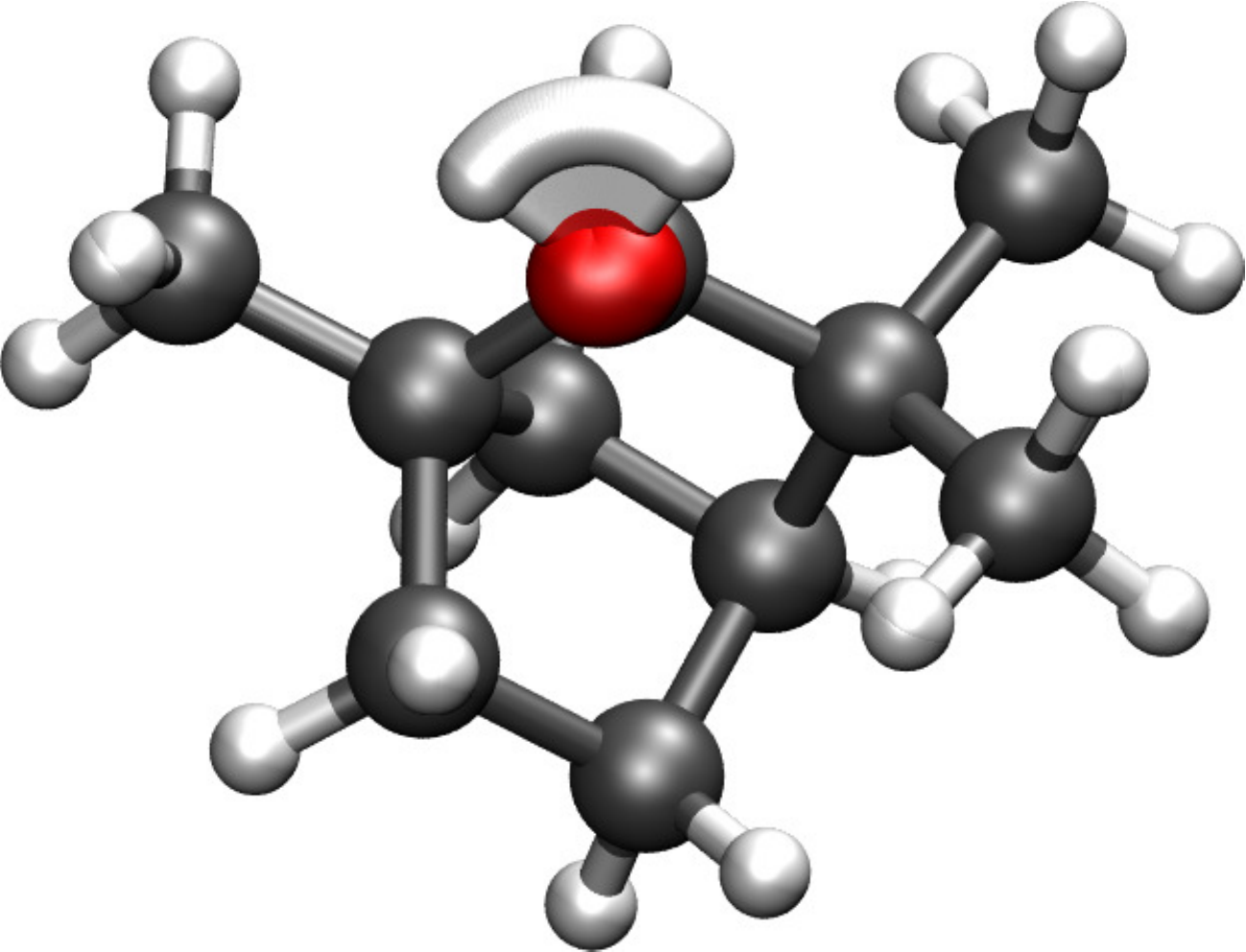}
    \caption{The instanton representation of the tunneling process in \textalpha-fenchol's H  isotopomer. }
    \label{fig:alpha-fenchol}
\end{figure}

A representation of the tunneling mechanism in \textalpha-fenchol's H isotopomer as determined by the instanton is shown in Fig.~\ref{fig:alpha-fenchol}.
Tunneling mainly occurs in the hydroxy group, 
with the H atom doing most of the tunneling (contributing 83\% to the action)
and a smaller but nevertheless significant contribution from the O atom (15\%).
For the D isotopomer, similar observations were made (85\% and 14\% for the D and O atom respectively).


\begin{figure}[h!]
    \centering
    \includegraphics[width=0.95\linewidth, keepaspectratio]{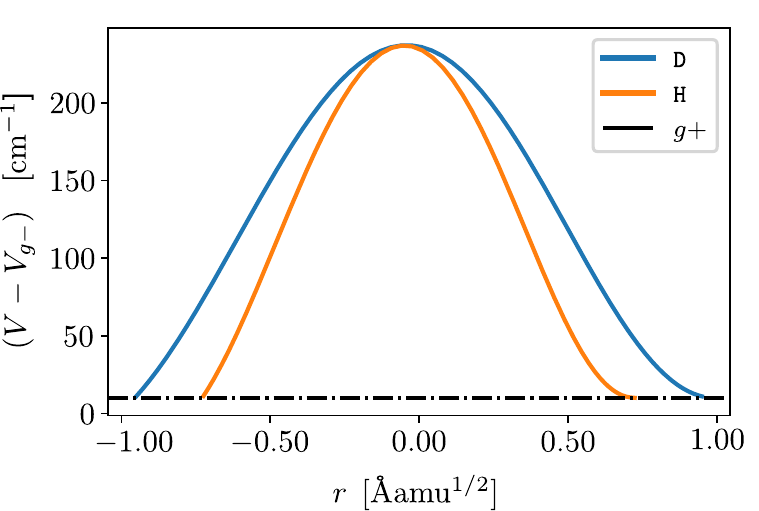}
    \caption{Potential energy along the mass-weighted instanton pathway for the H and D isotopomers of \textalpha-fenchol.
    The $g-$ minimum energy was set as the baseline, and the relative energy of the $g+$ minimum is indicated by the dashed line.}
    \label{fig:alpha-fenchol-instplot} 
\end{figure}

Fig.~\ref{fig:alpha-fenchol-instplot} presents a plot of the potential along the instanton trajectory as a function of cumulative mass-weighted path-length (defined in previous work\cite{GPR, HCH4}).
Similar to what has been observed in Sec.~\ref{subsec:2d},
the instanton trajectory starts (on the right) from the higher-energy minimum $g+$, crosses the barrier and
reaches a bounce point, 
but never reaches the $g-$ minimum.
One can also note that upon isotopic substitution,
the path length of the instanton increases,
which consequently means that the action is larger.
This will effectively reduce $\hbar\Omega$,
as we shall discuss later.


\begin{table*}[t]
\setlength{\tabcolsep}{8pt}
\begin{tabular}{llllll}
\hline\hline
Isotopomer         & Level of Theory        & Method            & $d$    & $\hbar \Omega$  & $\Delta$ \\ \hline
\multirow{4}{*}{H} & B3LYP                  & Instanton         & 12.6 & 10.4          &  32.9  \\ \cline{2-6}
                   & DLPNO-CCSD(T) // B3LYP & Instanton         & 4.25 & {7.60}    & {17.4}  \\
                   &                        & 1D approx\cite{fenchol} & 4.55 & 7.20           & 17.0  \\ \cline{2-6}
                   & -                      & Experiment \cite{fenchol}        & -    & -              & 16    \\ \hline\hline
\multirow{4}{*}{D} & B3LYP                  & Instanton         & 10.7 & 1.95           & 21.8      \\ \cline{2-6}
                   & DLPNO-CCSD(T) // B3LYP & Instanton         & 2.55 & {1.43}             & {5.85}       \\
                   &                        & 1D approx \cite{fenchol} & 2.30 & 1.30           & 5.30   \\ \cline{2-6}
                   & -                      & Experiment \cite{fenchol}        & -    & -              & 7    \\\hline\hline
\end{tabular}
\caption{Asymmetry contributions $d = \half\left({E}_r - {E}_\ell \right)$ and tunneling contributions $\hbar\Omega$ to the level splitting $\Delta = 2\sqrt{d^2 + \left(\hbar\Omega\right)^2}$,
all given in $\mathrm{cm^{-1}}$,
for the H and D isotopomer of \textalpha-fenchol.
}
\label{tbl:fenchol}
\end{table*}

The calculated level splittings as well as the experimental measurements are presented in Table \ref{tbl:fenchol}.
Here, we note that $d \simeq \hbar\Omega$,
which indicates significant contributions from tunneling as well as asymmetry.
The level splitting calculated for the H isotopomer with B3LYP-D3BJ/6-311++G(d,p) is approximately twice the experimental value;
however, this error decreases to only 9\% larger than the experimental value using the dual-level approach.
A similar trend is repeated for the D isotopomer,
wherein initially, 
the result obtained is significantly larger than the experimental value.
Upon correcting the electronic energies with DLNPO-CCSD(T),
the prediction is far more accurate,
with the discrepancy reduced such that the level splitting obtained with instanton theory is only 16\% smaller than the experimental value.

As expected, 
the level splitting of the D isotopomer is lower than that of the H isotopomer.
This is partly due to a change in the zero-point energy at the minimum,
which alters $d$.
The effect is however even stronger in the tunneling contribution $\hbar\Omega$,
where one can observe a decrease of approximately a factor of 6 upon isotopic substitution.
With instanton theory,
we can easily explain the decrease in $\hbar\Omega$ by attributing it to an increase in the mass-weighted path length,
which in turn increases the action as mentioned previously.

In Sec.~\ref{sec:theory},
we have defined a mixing angle $\phi$,\cite{asymtunnel}
which we use as a measure of delocalization.
The mixing angle is defined over the range $0^\circ\leq \phi \leq 180^\circ$;
close to $0^\circ$ or $180^\circ$,
the system's eigenstates $\psi_0$ and $\psi_1$ are almost completely localized on each well,
whereas if $\phi$ approaches $90^\circ$,
the system is maximally delocalized with half the population in each well.
The mixing angles we obtain for the H and D isotopomers are $60.7^\circ$ and $29.3^\circ$, respectively.
This suggests that the states are partially mixed,
which is in agreement with the conclusions of the experimental study.
One can also take note from these values that the D isotopomer is more localized than the H isotopomer.

Medel et al.\cite{fenchol}\ carried out a simple theoretical study wherein they modeled the rearrangement between the two conformers $g+$ and $g-$ as a simple torsional motion.
They then solved the resulting one-dimensional Schr\"odinger equation on a grid of dihedral angles in order to obtain the level splittings between the two lowest eigenstates.
Given that they also evaluated the energies of the $g+$ and $g-$ conformers,
they were easily able to deduce the values of $d$ and hence extract $\hbar\Omega$ from the level splitting.
The values they obtained are given in Table \ref{tbl:fenchol}, and are similar to ours;
the minor discrepancies in $d$ could be attributed to the difference in basis set employed (we used the \text{6-311G++(d,p)} basis set, while they chose may-cc-pVTZ).
In a sense, it is surprising that the simple one-dimensional approach gave such good agreement with experiment; not only does it mean that chemical intuition was sufficient for guessing the right tunneling path (meaning that there is little corner cutting), it also implies the change in the frequencies of perpendicular degrees of freedom along the path is negligible in this case. This is certainly not true in general, as previous work has demonstrated the importance of a full-dimensional calculation. \cite{formic}
Instanton theory does not rely on these assumptions, making it a much more robust approach for calculating the tunneling splitting. 

\section{Connection to rate theory} \label{subsec:discuss-rate}

The instanton pathway used in this work to calculate the tunneling splitting is identical to the ``bounce'' path considered by Coleman in his analysis of the ``fate of the false vacuum''. \cite{Coleman1977ImF,Uses_of_Instantons}
The same instanton is also used to calculate unimolecular tunneling rates in the low-temperature limit. \cite{Miller1975semiclassical,Benderskii,Weiss,Kaestner2013carbenes,Zhou2023peroxide,carbenes}
This similarity may seem surprising, as tunneling splittings and reaction rates pertain to fundamentally different scenarios: tunneling splittings are relevant for symmetric or almost symmetric double-well systems where the product state is bounded, and are associated with coherent oscillatory dynamics; rates on the other hand are relevant in the case of metastable wells and are associated with an incoherent exponential decay. However, the instanton cannot distinguish between these two scenarios; it only probes the barrier region of the system and does not actually descend into the lower-energy product minimum, meaning it has no information on whether the  product state is bounded or an open scattering channel. This explains how the same instanton can be used for the calculation of both the low-temperature rate constant and the tunneling splitting. 



This apparent connection between the low-temperature rate constant and tunneling splitting in instanton theory motivates us to find a mathematical relationship between the two. We note that some analysis in this direction has already been carried out by Miller\cite{Miller1979periodic} based on a WKB formalism. 
He derived an expression for the change in the level splitting due to tunneling in the strongly asymmetric regime, where $\Delta/2=d\sqrt{1+\Omega^2/d^2}\approx d +\Omega^2/2d$, meaning the energy shift of a level due to tunneling is given by $\Omega^2/2d$. Comparing it to a semiclassical expression for the decay rate of a metastable state, he found that the rate is proportional to this energy shift, i.e.\ it scales as $k\propto \Omega^2$.
This is in contrast to another widespread formula \cite{sedgi2023playground,christoffel1981quantum,bowman1986self,bosch1990bidimensional,Mediavilla2008cyclobutadiene}, $k=2\Omega/\pi$, which is based on the short-time limit of the coherent dynamics of a wavepacket in a double well.\cite{townesmicrowave,brickmann1969lingering}
Our discussion below corroborates Miller's result, and thereby indicates that the  relation $k=2\Omega/\pi$ is incorrect, at least in the context which we discuss. This is also in agreement with the results of a recent empirical study,\cite{rodriguez2025tunneling} in which the $k\propto \Omega^2$ relationship was observed in a comparison between computed rate constants and experimentally measured tunneling splittings for a range of compounds.  

For a reaction from right to left (i.e.\ higher to lower well), the rate constant is defined by \cite{QInst}
\begin{align}\label{eq:QInst}
    k_{r\rightarrow\ell} Z_r(\beta\hbar) &\simeq \int_{-\infty}^\infty c_\mathrm{ff}(\tau+\iu t) \, \rmd t \nonumber
     \\&\simeq c_\mathrm{ff}(\tau) \sqrt\frac{2\pi}{\phi''} ,
\end{align}
where one should keep in mind that $c_\mathrm{ff}(\tau+\iu t)$ is the projected flux correlation function; semiclassically, its time integral is only half that of the standard flux correlation function.\cite{Miller1983rate} To reach the second line,
we employ the steepest-descent approximation. We start by writing 
$c_\mathrm{ff}(\tau+\iu t) 
= c_\mathrm{ff}(\tau) \eu{-\phi}$
with
$\phi = \ln c_\mathrm{ff}(\tau) - \ln c_\mathrm{ff}(\tau+\iu t)$.
We then Taylor expand $\phi$ to second order around $t=0$ such that  
$\phi \approx \half \phi'' t^2$, where $\phi'=0$ due to our appropriate choice of $\tau$ and $\phi'' \simeq \frac{1}{\hbar}\left(\pder{E_\ell}{\tau_\ell}+\pder{E_r}{\tau_r}\right)$ within a semiclassical approximation.\cite{QInst}
This can be simplified further by noting that in the limit of large $\beta$,
$\frac{1}{\hbar}\pder{E_r}{\tau_r} \rightarrow 0$ such that $\phi'' \approx \frac{1}{\hbar}\pder{E_\ell}{\tau_\ell}$.

Applying the expression we obtained for $c_\mathrm{ff}(\tau)$ in this paper [Eq.~(\ref{CffPQM})],
we obtain
\begin{align}
    k_{r\rightarrow \ell} Z_r(\beta\hbar) &\simeq \Omega^2 Z_\ell(\tau) Z_r(\beta\hbar-\tau)\sqrt{\frac{2\pi}{\phi''}} \nonumber \\
    &\simeq \Omega^2 Z_r(\beta\hbar)   \sqrt{\frac{2\pi}{\phi''}} \, \eu{2\tau d / \hbar}.
\end{align}
The relationship between the tunneling frequency $\Omega$,
and the rate constant $k_{r\rightarrow\ell} $ is therefore
\begin{align} \label{eq:kOmega2}
    k_{r\rightarrow\ell} \simeq \Omega^2  \sqrt{2\pi\hbar} \left(\pder{E_\ell}{\tau_\ell}\right)^{-\half} \eu{2\tau d / \hbar}.
\end{align}
Thus, as was also found by Miller through a different argument, \cite{Miller1979periodic}
there is a quadratic dependence of the rate on the tunneling matrix element.
For a symmetric double-well system, the low-temperature limit of the rate constant becomes undefined (as $\pder{E_\ell}{\tau_\ell}\rightarrow0$), which is not unexpected as the quantum-mechanical rate is also undefined in this case.\cite{Weiss} 

To demonstrate this relationship numerically, we design a new one-dimensional model system. For the instanton rate constant to be well-defined, the well depths of the double well need to be sufficiently asymmetric, while for the tunneling matrix element to be meaningful the two lowest-lying states should be much closer in energy than the vibrational energy-level spacing. Additionally, to be able to calculate a well-defined quantum-mechanical rate constant, it should be possible to turn the double well into a metastable well. 

\begin{figure}[h]
    \centering
    \includegraphics[width=0.7\linewidth]{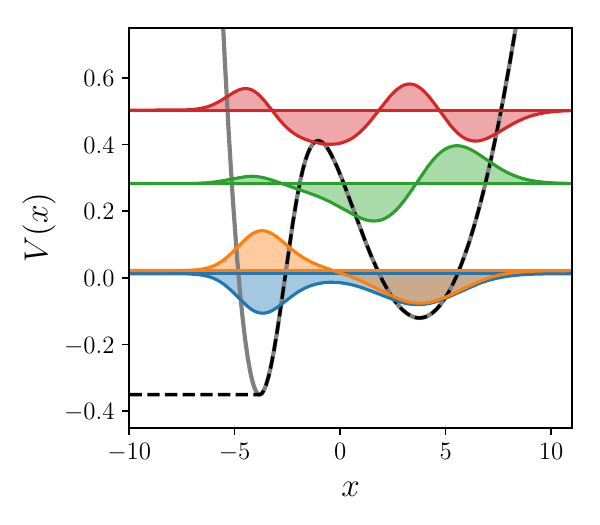}
    \caption{Model potential for comparing rate constants to tunneling splittings. The double-well potential (gray) and its four lowest eigenstates are shown, along with the associated metastable well (black dashes) for $x_0=4.1$. }
    \label{fig:komegaV} 
\end{figure}

A potential that fulfils these requirements is given by
\begin{align} \label{eq:komega_pot}
    V(x) = \half(V_{00}+V_{11})-\half\sqrt{(V_{00}-V_{11})^2+4 V_{01}^2}
\end{align}
where $V_{00}=\half\omega_\ell^2(x+x_0)^2-\varepsilon$ and
$V_{11}=\half\omega_r^2(x-x_0)^2$.
We choose $\omega_r=0.3$, $\omega_\ell=1.0$ and $V_{01}=2.0$ in reduced units where $m=\hbar=1$. By changing 
$x_0$ we can modify the barrier height. We tune the bias $\varepsilon$ such that the lowest two eigenstates remain close enough in energy for the two-level system description to be meaningful, as shown in Fig.~\ref{fig:komegaV}. In particular, we tune it such as to minimize the level splitting $\Delta$ obtained from a DVR calculation and extract the tunneling frequency as $\hbar\Omega_\text{QM} = \Delta_\mathrm{QM}/2$. 

\begin{table*}[t]
\centering
\setlength{\tabcolsep}{6pt} 
\begin{tabular}{llllllll} 
\hline\hline
$x_0$  & barrier height & $\Omega_\mathrm{inst}   $  &   $\Omega_\mathrm{QM}   $     & Error in $\Omega$ [\%] & $k_\mathrm{inst}$   & $k_\mathrm{QM}$      & Error in $k$ (\%) \\
\hline
4.1 & 0.53       & 6.31(--3) & 4.58(--3) & 38         &  1.74(--4) & 2.13(--4) & --19        \\
5   & 1.26        & 1.12(--4) & 9.22(--5) & 21         &  5.59(--8) & 7.00(--8) & --20        \\
5.6 & 1.98        & 2.97(--6) & 2.51(--6) & 18         &  3.95(--11) & 4.8(--11)  & --18    \\
\hline \hline
\end{tabular}
\caption{Comparison of the tunneling frequency and the low-temperature limit of the rate constant, as calculated with instanton theory and with exact methods.
We consider three different barrier heights (tuned by the parameter $x_0$ with $\varepsilon$ defined to minimize $\Delta_\mathrm{QM}$ in each case), for the potential given by Eq.~\eqref{eq:komega_pot}.
}\label{tab:komega}
\end{table*}

For calculating the quantum-mechanical rate, we convert the double well to a metastable well by flattening the potential to the left of the lower minimum well (black dashes in Fig.~\ref{fig:komegaV}). We then construct scattering wavefunctions using the Numerov method,\cite{Allison1970numerov} from which we extract the scattering phase shift as a function of energy.\cite{Child} Around a resonance, the phase shift changes by $\pi$ and the resonance peak is described by a Lorentzian to an excellent approximation; the width of this resonance feature gives the rate of escape $\Gamma_j$ from the corresponding $j$th metastable state of the right well. We can relate this to a thermal rate constant by realizing that in the low-temperature limit, only the lowest-lying metastable state will be occupied and contribute to the rate, so that $k=\Gamma_0$.

The instanton rate can be calculated either directly within standard ring-polymer instanton rate theory,\cite{InstReview} or via Eq.~\eqref{eq:kOmega2}, using $\Omega=\Omega_\mathrm{inst}$, $d=V_r + \tfrac{1}{2}\hbar\omega_r - V_\ell - \tfrac{1}{2}\hbar \omega_\ell$ and obtaining $\pder{E_\ell}{\tau_\ell}=\tfrac{\partial^2 S_\ell}{\partial \tau_\ell^2}$ in this one-dimensional system 
as $\pder{E_\ell}{\tau_\ell} = - (\tfrac{\partial^2 W_\ell}{\partial E^2})^{-1}$.\cite{GoldenGreens,GoldenRPI,QInst} Both approaches yield the same result. 

The results for three different barrier heights are shown in Table \ref{tab:komega}. In all three cases, the instanton results for both the tunneling splitting and the rate are in good agreement with the quantum results, consolidating the validity of the semiclassical relation we established between the two.

\section{Conclusions} \label{sec:conclusions}
In this work, we have extended instanton theory to describe tunneling splittings in asymmetric systems. This new formulation, based on the projected flux correlation function, is more general than our previous asymmetric version of instanton theory. \cite{asymtunnel} In particular, not only can it describe tunneling between wells of unequal frequency, it is also applicable to wells of unequal depth. 

We note that an asymmetric version of Jacobi-fields instanton theory has recently been proposed by Erakovi\'c and Cvita\v{s} based on a generalization of the Herring formula.\cite{Erakovic2022instanton}
It is hard to directly compare their method with ours as the underlying concepts are so disparate.
However, it is clear that the two methods must be subtly different, as our approach remains a periodic orbit, whereas the Jacobi-fields instanton has a discontinuity in the energy halfway along the path such that it can reach the bottom of both wells.
Nonetheless, in the case of weak asymmetry there is no reason to expect a large difference in the predictions.
It is left for future work to determine if there are cases where one method has a particular advantage over the other.
It will be hard to argue from the perspective of asymptotic analysis,\cite{BenderBook} as in this case, the level splitting will always be dominated by $d$ in the $\hbar\rightarrow0$ limit.
One practical advantage of our approach is that the tunneling pathway can be optimized with no changes to existing ring-polymer instanton code.\cite{InstReview,molpro,iPI2}

We have tested our new method on 
one- and two-dimensional asymmetric model systems and found that the instanton results are in very good agreement with the quantum-mechanical benchmark. We have furthermore applied it to calculate the level splitting in full dimensionality in  \textalpha-fenchol, an asymmetric biomolecule for which experimental values for the level splitting are available. In order to make instanton calculations feasible for this relatively large molecule, we used machine learning to significantly reduce the number of electronic-structure calculations required. The resulting level splittings obtained with instanton theory are within 20\% of the experimental values. Although in this case, a similar result was attained using a 1D model,
our method includes multidimensional effects (which appear to cancel out in this particular case) and avoids the difficulty of having to \emph{a priori} guess the tunneling coordinate, making it a more much reliable approach in general.

Interestingly, the similarity between our newly developed instanton theory for tunneling splittings and instanton rate theory allows for comparison between the two; from this we have obtained a relation between the rate $k$ and tunneling frequency $\Omega$. In particular, we confirm Miller's claim \cite{Miller1979periodic} that $k\propto \Omega^2$, in contrast to the often suggested linear dependence. 

In addition to describing asymmetric molecules,
other potential applications of our new method include the simulation of tunneling 
in asymmetric environments.
An example of this would be the water hexamer cage.\cite{Gregory1997hexamer,Perez2012hexamer,Walesdimerhexamer} 
Another example is 
the study of glasses: their energy landscapes have many minima, and some of those may be close in energy and separated only by a low barrier, thereby giving rise to asymmetric double-well systems in which tunneling plays a prominent role. Physically, transitions between these wells can range from single atom tunneling to rearrangements involving tens to hundreds of atoms. \cite{reinisch2005moving,rubio2014identification,damart2018atomistic,muller2019towards,khomenko2020depletion,berthier2023modern} Of particular interest are double-well systems with a level splitting on the order of 1 K, as these are hypothesized to be origin of the anomalous linear (rather than cubic) temperature dependence of the heat capacity at low temperature that glasses are known for. \cite{zeller1971thermal,anderson1971anomalous,phillips1972tunneling} The prefactor of this linear heat capacity depends on $n(\Delta)$, the cumulative density of double-well systems with level splitting $\Delta$; a reliable prediction of this quantity requires an accurate method for calculating $\Delta$. Previous studies have resorted to \emph{e.g.}\ the WKB approximation \cite{reinisch2005moving} or reduced-dimensional approximations,\cite{khomenko2020depletion} 
but instanton theory applied to the full-dimensional system should provide a more accurate approach. This may help close the gap between theory and experiment in this field, as the discrepancy in $n(\Delta)$ is currently 2 to 3 orders of magnitude.\cite{khomenko2020depletion}

In future work we aim to build upon this method in order to improve the description of tunneling currently provided by our theory.
The small discrepancies between instanton theory and numerically-exact quantum mechanics could be addressed through pertubative corrections,
as an extension of our theory for the symmetric-tunneling case \cite{AnharmInst} and rate theory.\cite{PCIRT}
Following Erakovi\'{c} and Cvita\v{s}, the accuracy of the asymmetry contribution $d$ could be improved through vibrational configuration interaction calculations. \cite{Cvitas2022VCI}
In a forthcoming paper we shall further extend our theory to account for vibrationally-excited states.
The information provided by splittings of vibrationally-excited states could prove very insightful;
it can for example allow us to quantify whether certain vibrational states contribute or suppress tunneling. We find that this theory is also naturally expressed in terms of projected flux correlation functions, making the theory presented in this work of deeper significance.

\section*{Acknowledgements}
The authors acknowledge financial support from the Swiss National Science Foundation through Project 175696.
M.R.F.\ was supported by an
ETH Zurich Research Grant.

\section*{Data Availability}
The data that supports the findings of this study are available within the article.

\bibliography{references,extra,marit}

\end{document}